\documentclass[10pt]{article}
\usepackage{jheppub}

\newcommand{\ba}{\begin{eqnarray}}
\newcommand{\ea}{\end{eqnarray}}
\newcommand{\af}{a_f}
\newcommand{\cf}{c_f}
\newcommand{\nf}{n_f}
\newcommand{\str}{{\rm sTr}}
\newcommand{\la}[1]{\label{#1}}

\newcommand{\eq}{eq.~}
\newcommand{\se}{section~}
\newcommand{\secs}{sections~}

\newcommand{\eqs}{eqs.~}
\newcommand{\nr}[1]{(\ref{#1})}
\newcommand{\ep}{\varepsilon}
\newcommand{\CA}{C_{\mathrm{A}}}
\newcommand{\CF}{C_{\mathrm{F}}}
\newcommand{\NA}{N_{\mathrm{A}}}
\newcommand{\NF}{N_{\mathrm{F}}}
\newcommand{\TF}{T_{\mathrm{F}}}
\newcommand{\Nf}{N_{\mathrm{f}}}
\newcommand{\order}[1]{{\cal O}(#1)}
\newcommand{\msbar}{{\overline{\mbox{\rm{MS}}}}}
\newcommand{\code}[1]{{\tt #1}}

\newcommand{\gammaGG}{\gamma_3}
\newcommand{\gammaGGG}{\gamma_1^{3g}}
\newcommand{\gammaGGGG}{\gamma_1^{4g}}
\newcommand{\gammaCC}{\gamma_3^{c}}
\newcommand{\gammaCc}[1]{\gamma_{3#1}^{c}}
\newcommand{\gammaCCG}{\gamma_1^{ccg}}
\newcommand{\gammaCcg}[1]{\gamma_{1#1}^{ccg}}
\newcommand{\gammaQQ}{\gamma_2}
\newcommand{\gammaQQG}{\gamma_1^{\psi\psi g}}

\title{Complete renormalization of QCD at five loops}

\preprint{\mbox{}\hfill BI-TP 2017/01\\\mbox{}\hfill DESY 17-011\\\mbox{}\hfill IPPP/17/7}

\author[a]{Thomas Luthe,}
\author[b]{Andreas Maier,}
\author[c]{Peter Marquard}
\author[d]{and York Schr\"oder}

\affiliation[a]{Faculty of Physics, University of Bielefeld, 33501 Bielefeld, Germany}
\affiliation[b]{Institute for Particle Physics Phenomenology, Durham University, Durham, United Kingdom}
\affiliation[c]{Deutsches Elektronen Synchrotron (DESY), Platanenallee 6, Zeuthen, Germany}
\affiliation[d]{Grupo de F\'isica de Altas Energ\'ias, Universidad del B\'io-B\'io, Casilla 447, Chill\'an, Chile}

\emailAdd{tluthe@physik.uni-bielefeld.de}
\emailAdd{andreas.maier@durham.ac.uk}
\emailAdd{peter.marquard@desy.de}
\emailAdd{yschroder@ubiobio.cl}


\abstract{We present new analytical five-loop Feynman-gauge results for the anomalous dimensions of ghost field and -vertex, generalizing the known values for SU(3) to a general gauge group. Together with previously published results on the quark mass and -field anomalous dimensions and the Beta function, this completes the 5-loop renormalization program of gauge theories in that gauge.}

\begin{document}
\maketitle

%
\section{Introduction}
\la{se:intro}

Non-Abelian quantum field theories have become a cornerstone of our description of nature, ever since it has been understood that they allow for the interaction strength to become weaker as the energy rises, a property called 'asymptotic freedom' \cite{Gross:1973id,Politzer:1973fx,thooft}. Providing for an explanation of quark confinement and hence a successful theory of the strong interactions that bind together quarks into protons, Quantum Chromodynamics (QCD), much theoretical work has been focused on this central part of the Standard Model. The change (or 'running') of its couplings and masses with energy is governed by renormalization group equations, which in turn need the theory's anomalous dimensions as an input. In the light of ever-increasing precision of collider experiments conducted at various energies, a first-principles determination of these parameters with the highest achievable theoretical accuracy has become necessary. 

One of the hallmarks of perturbative renormalization is the so-called Beta function, governing the scale dependence of the renormalized strong coupling constant. For scalar $\phi^4$ theory, for example, renormalization has recently been pushed to the 6-loop level \cite{Batkovich:2016jus,Kompaniets:2016hct}. For the case of QCD, the current precision frontier is at five loops \cite{Baikov:2016tgj}, and most recently, results for a generalization from the gauge group SU(3) to general simple Lie groups have appeared \cite{Luthe:2016ima,Herzog:2017ohr}. The Beta function represents one of the above-mentioned anomalous dimensions that are contained in the renormalization group equations. In total there are five linearly independent ones, such as the quark mass anomalous dimension (for 5-loop results, see \cite{Baikov:2014qja,Luthe:2016xec}) that is of key phenomenological importance when evolving the masses typically measured at a few GeV to the electroweak scale, as well as three further non-physical coefficients that we will discuss in the following section.

In the present paper, our aim is to push the knowledge of the remaining renormalization constants to the same level. We follow up on our earlier works \cite{Luthe:2016ima,Luthe:2016xec} and complete the renormalization program by evaluating the anomalous dimension of the ghost field, and that of the ghost-gluon vertex. Using the same general techniques as in our previous works, we have been able to compute these coefficients in Feynman gauge (although a result for general covariant gauges is conceptionally not more involved, and could be obtained in a straightforward way given more computational resources).

The structure of the paper is as follows. After briefly summarizing our setup and introducing some notation in \secs\ref{se:setup}, we give our results for the anomalous dimensions of the ghost sector in \ref{se:ghost}. In \se\ref{se:rest} we display the relations needed to obtain the remaining anomalous dimensions, and to derive all corresponding renormalization constants. All these results are lengthy and given in ancillary files for convenience. We conclude in \se\ref{se:conclu}. In the appendix, the known Beta function coefficients are summarized in our notation.

%
\section{Setup}
\la{se:setup}

In this section, we start by fixing some necessary conventions and notation, and comment on the computational strategy that we employ for our five-loop computation.

%
\subsection{Renormalization constants}

The fields and parameters of the gauge theory are renormalized via 
\ba
\psi_b &=& \sqrt{Z_2}\psi_r \;,\quad A_b = \sqrt{Z_3}A_r \;,\quad c_b = \sqrt{Z_3^c}c_r \;,\\
\la{eq:ZgDef}
m_b &=& Z_m m_r \;,\quad g_b = \mu^\ep Z_g g_r \;,\quad \xi_{L,b} = Z_\xi \xi_{L,r} \;,
\ea
where the subscript $b$ ($r$) stands for bare (renormalized) quantities, and all $Z_i=1+{\cal O}(g_r^2)$. There is no need to renormalize the gauge-fixing term $\sim(\partial A)^2/\xi_L$, such that setting $Z_\xi=Z_3$ leaves five independent renormalization constants (RCs) $Z_i$.
It is sometimes convenient to consider products of the $Z_i$ as 'vertex RCs', such as those multiplying the 3-gluon, 4-gluon, ghost-gluon and quark-gluon vertex. These are usually denoted as $Z_1^j$, where $j\in\{3g,4g,ccg,\psi\psi g\}$, and we will find it convenient to evaluate the combination $Z_1^{ccg}=\sqrt{Z_3}\,Z_3^c\,Z_g$ instead of $Z_g$.
For a complete set of relations among these constants, see \se\ref{se:rest} below.

Instead of explicitly listing the renormalization constants $Z_i$, we will for simplicity only give the corresponding anomalous dimensions, defined by 
\ba
\la{eq:gammaDef}
\gamma_i &=& -\partial_{\ln\mu^2}\ln Z_i \;.
\ea
Two of them, $\gamma_m$ and $\gamma_2$, corresponding to the renormalization of the quark mass and wave function, have already been given in \cite{Luthe:2016xec}. Three more are required, in order to complete the set of independent RCs. A popular choice is to evaluate $\gamma_3$, $\gamma_3^c$ and $\gamma_1^{ccg}$, the latter two of which we provide in the present paper, while $\gamma_3$ can then be reconstructed from the Beta function given in \cite{Herzog:2017ohr}, see \eq\nr{eq:gammaGG} below. 

Note that $\gamma_3$ is conveniently traded for the Beta function, since the latter is a physical gauge invariant object and hence much more compact. Following usual conventions, instead of considering $Z_g$, one renormalizes the gauge coupling squared (which in our notation is $a\sim g^2$, see \eq\nr{eq:a} below) with the factor $Z_a\equiv Z_g^{\,2}$ and calls the corresponding anomalous dimension the Beta function, $\beta\equiv\gamma_a=2\gamma_g$. Note that, due to the renormalization scale independence of the bare gauge coupling, using \eqs\nr{eq:ZgDef} and \nr{eq:gammaDef} this immediately implies
\ba
\label{eq:betaDef}
\beta=\ep+\partial_{\ln\mu^2}\ln a \quad\Leftrightarrow\quad \partial_{\ln\mu^2}a = -a\Big[\ep-\beta\Big]\;,
\ea
where $a$ is the renormalized coupling of \eq\nr{eq:a}.

%
\subsection{Computational strategy}

In order to evaluate the coefficients of the perturbative expansions for the renormalization constants and anomalous dimensions, we rely on the setup that we have developed and successfully tested in our previous works. While we refer to the literature for more detailed descriptions of our chain of computer-algebra programs \cite{Luthe:2016ima,Luthe:2016xec} and for technical details concerning our procedure for integral reduction and evaluation \cite{crusher,Luthe:2015ngq,Luthe:2016sya,Luthe:2016spi}, let us briefly summarize the main components of our strategy here. 

As is standard procedure in perturbative multiloop calculations, we start with expressions produced by the diagram generator \code{qgraf} \cite{Nogueira:1991ex,Nogueira:2006pq} linked with some own \code{FORM} \cite{Vermaseren:2000nd,Tentyukov:2007mu,Kuipers:2012rf} codes.
Taking the required set of 2- and 3-point functions, we perform the group algebra with the help of \code{color} \cite{vanRitbergen:1998pn}. 
Next, we introduce a common mass term into all our massless propagators \cite{Misiak:1994zw,vanRitbergen:1997va,Chetyrkin:1997fm}, in order to regulate the infrared behavior of the dimensionally regularized \cite{Bollini:1972ui,tHooft:1972tcz} $d=4-2\ep$ dimensional momentum-space integrals. While this does not change the ultraviolet (UV) behavior that we are interested in when extracting the (mass-independent) UV counterterms in the $\msbar$ scheme \cite{tHooft:1973mfk}, it allows us to perform a systematic expansion in external momenta, which can eventually be nullified. This leaves us with expansions coefficients that belong to the well-studied class of fully massive vacuum integrals.

At five loops, this class of integrals can be labelled by 15 indices (corresponding to maximally 12 propagators plus 3 scalar products) \cite{Luthe:2015ngq}. To tame the enormous number of integrals that enter our calculation, we choose to perform a reduction to a small set of master integrals. To this end, we make use of our own codes \code{crusher} \cite{crusher} and \code{TIDE} \cite{Luthe:2015ngq}, which are based on integration-by-parts (IBP) identities \cite{Chetyrkin:1981qh} and use Laporta-type algorithms \cite{Laporta:2001dd}. Both \code{C++} codes are largely independent, utilize \code{GiNaC} \cite{Bauer:2000cp} and \code{Fermat} \cite{fermat} for simple and complicated algebraic manipulations, respectively, and in conjunction provide us with a welcome verification of the time- and resource-consuming integral reduction process.

After reduction, we end up with a set of 110 five-loop master integrals. Their high-precision numerical $\ep$\/-expansion has been studied previously \cite{Luthe:2015ngq,Luthe:2016sya}, much along the lines of previous work on the four-loop case \cite{vanRitbergen:1997va,Laporta:2002pg,Schroder:2002re,Czakon:2004bu,Schroder:2005va}, relying again on IBP reductions, generating large coupled systems of linear difference equations that can be solved formally with factorial series \cite{Laporta:2001dd}. A truncated version of these series then delivers high-precision numerical results for the coefficients of the $\ep$\/-expansion of each individual master integral. This allows to employ the integer-relation finding algorithm \code{PSLQ} \cite{MR1489971}, testing for relations between some of these numbers, and discovering the analytic content of others. 

Our high-precision evaluation of all 5-loop master integrals has not yet produced results for the 12-line families \cite{Luthe:2016ima}. Fortunately, it turns out that in all our results, only three independent linear combinations of 12-line master integrals contribute. While standard integration over the Feynman parametric representation 
(see e.g.\ \cite{Smirnov:2012gma}) with subsequent sector decomposition, using the strategy explained in \cite{Roth:1996pd,Binoth:2003ak} and as implemented in \code{FIESTA} \cite{Smirnov:2015mct} as well as own code \cite{Luthe:2016spi} gives 3-6 digits only, we were able to fix the three unknown linear combinations to 260 digits, as explained in \cite{Luthe:2016xec}. Owing to this last step, we are able to provide analytic expressions for the specific combinations of master integrals that appear in all of our results below.

%
\subsection{Notation for group invariants}
\la{se:notation}

In order to render the present paper self-contained, we wish to recall some group-theoretic notation that we had already utilized in our previous works \cite{Luthe:2016ima,Luthe:2016xec}, and which we employ to present all of our results below. We study a Yang-Mills theory coupled to fermions, working over a semi-simple Lie algebra with hermitian generators $T^a$.  The commutation relations $T^aT^b-T^bT^a=i f^{abc}T^c$ define the real and antisymmetric structure constants $f^{abc}$.
Following standard conventions, the quadratic Casimir operators of the fundamental and adjoint representations (of dimensions $\NF$ and $\NA$, respectively) are defined via $T^aT^a=\CF1\!\!1$ and $f^{acd}f^{bcd}=\CA\delta^{ab}$.
The trace normalization reads ${\rm Tr}(T^aT^b)=\TF\delta^{ab}$, the number of quark flavors is denoted by $\Nf$, and we will make use of the following normalized combinations of group invariants:
\ba
\nf=\frac{\Nf\,\TF}{\CA} \quad,\quad 
\cf=\frac{\CF}{\CA} \;.
\ea

In loop diagrams, higher-order group invariants arise when one encounters traces of more than two group generators. It is useful to classify these higher-order traces in terms of combinations of symmetric tensors \cite{vanRitbergen:1998pn}, of which we presently need the following three (rewriting the generators of the adjoint representation as $[F^a]_{bc}=-if^{abc}$, and again normalizing conveniently):
\ba
d_1=\frac{[\str(T^aT^bT^cT^d)]^2}{\NA\TF^2\CA^2} \;,\;
d_2=\frac{\str(T^aT^bT^cT^d)\,\str(F^aF^bF^cF^d)}{\NA\TF\CA^3} \;,\;
d_3=\frac{[\str(F^aF^bF^cF^d)]^2}{\NA\CA^4} \;.
\ea
In the above, $\str$ denotes a fully symmetrized trace (such that $\str(ABC)=\frac12{\rm Tr}(ABC+ACB)$ etc.).

As a concrete example, picking SU($N$) as gauge group (and setting $\TF=\frac12$ and $\CA=N$), our set of normalized invariants reads \cite{vanRitbergen:1998pn}
\ba
\label{eq:sun}
\mbox{SU}(N):\;\;
\nf=\frac{\Nf}{2N}
\;,\;\;
\cf=\frac{N^2-1}{2N^2}
\;,\;\;
d_1=\frac{N^4-6N^2+18}{24N^4}
\;,\;\;
d_2=\frac{N^2+6}{24N^2}
\;,\;\;
d_3=\frac{N^2+36}{24N^2}\;.\quad
\ea
In the case of SU(3) (corresponding to physical QCD), we therefore have
\ba
\label{eq:su3}
\mbox{SU}(3):\;\;
\nf=\frac{\Nf}{6}
\;,\quad
\cf=\frac{4}{9}
\;,\quad
d_1=\frac{5}{216}
\;,\quad
d_2=\frac{5}{72}
\;,\quad
d_3=\frac{5}{24}
\;.\hspace*{43mm}
\ea
Results for the group U(1) (corresponding to QED) can be obtained by setting
\ba
\label{eq:u1}
\mbox{U}(1):\;\;
\CA=0 \;,\;\;
\CF=1 \;,\;\;
\TF=1 \;,\;\;
\NA=1 \;,\;\;
\str(T^aT^bT^cT^d)=1\;,\;\;
\str(F^aF^bF^cF^d)=0 \;,
\quad
\ea
which, due to our normalization, is sometimes only possible after multiplying with the corresponding power of the gauge coupling (that we normalize with a positive power of $\CA$, see \eq\nr{eq:a} below), eliminating all inverse powers of $\CA$.

%
\section{Renormalization of ghost field and -vertex}
\label{se:ghost}

In this section, we present our new results for the ghost field and ghost-gluon vertex anomalous dimensions at five loops. Since these are gauge-dependent quantities, let us stress once more that we have worked in Feynman gauge only. For lower loops, we display the full gauge parameter dependence, where $\xi=0/1$ corresponds to Feynman/Landau gauge.

%
\subsection{Ghost field anomalous dimension}

In contrast to the physical and gauge-independent Beta function and quark mass anomalous dimension, the anomalous dimension of the ghost field depends on the gauge parameter $\xi$. Its structure is
\ba
\label{eq:123loop}
\gammaCC &=& -a \Big[ -\tfrac14(2 + \xi)
+ \gammaCc{1} a+ \gammaCc{2} a^2 + \gammaCc{3} a^3+ \gammaCc{4} a^4
+ \dots\Big] 
\;,\quad
\label{eq:a}
a\equiv\frac{\CA\,g^2(\mu)}{16\pi^2}
\;,
\ea
where $g(\mu)$ is the gauge coupling constant that depends on the renormalization scale $\mu$, and we work in the $\msbar$ scheme, in $d=4-2\ep$ space-time dimensions. The 2- and 3-loop coefficients are known to be (see, e.g.\ \cite{Chetyrkin:2004mf})
\ba
\label{eq:2loop}
2^5\,3^1\,\gammaCc{1} &=& 5[16 \nf] - 2 (98 - 3 \xi) \;,\\
\label{eq:3loop}
2^8\,3^3\,\gammaCc{2} &=& 35[16 \nf]^2 
+ \big(324 (15 - 16 \zeta_3)\cf + 2 (5 + 189 \xi + 1944 \zeta_3)\big) [16 \nf] 
\nonumber\\&&
- 4 (14656 + 1485 \xi - 405 \xi^2 + 81 \xi^3) - 648 (4 - \xi) (2 - \xi) \zeta_3 \;.
\ea
The 4-loop coefficient $\gammaCc{3}$ is known for the gauge group SU($N$) \cite{Chetyrkin:2004mf}, while for a general Lie group the result is only available up to the linear term in an expansion around Feynman gauge $\xi=0$ \cite{Czakon:2004bu}. Unfortunately, due to the SU($N$) degeneracies $2d_2=7/12-\cf$ and $d_3=37/24-3\cf$, it is not possible to uniquely reconstruct its remaining gauge dependence (up to three loops the reconstruction works, since only quadratic Casimir operators contribute). We have therefore computed $\gammaCc{3}$ in general covariant gauge from scratch, obtaining (to clearly expose the group structure of the coefficients, we employ a notation resembling scalar products with vectors in curly brackets, such as e.g.\ $\{\cf,1\}.\{a,b\}=\cf a+b$)
\ba
\label{eq:4loop}
2^{11}3^4 \gammaCc{3} &=& 
(83 - 144 \zeta_3)[16\nf]^3 
\nonumber\\&&
+\Big\{\cf, 1\Big\}.\Big\{24 (1080 \zeta_3 \! -\! 648 \zeta_4 \!-\!  115), 
    2 (779 \xi \!-\!  8315)/3 - 432 (43 \!+\!  2 \xi) \zeta_3 + 11664 \zeta_4\Big\}[16\nf]^2  
\nonumber\\&&
+\Big\{\cf^2, d_2, \cf, 1\Big\}.\Big\{-864 (271 + 888 \zeta_3 - 1440 \zeta_5), 
 124416 (4 \zeta_3 - 5 \zeta_5), 
\nonumber\\&&
 24 \big(22517 + 3825 \xi - 864 (43 + \xi) \zeta_3 + 1296 (23 - \xi) \zeta_4 - 25920 \zeta_5\big),
 432 (2983 + 42 \xi - 6 \xi^2) \zeta_3 
\nonumber\\&&
- 648 (846 - 46 \xi + \xi^2) \zeta_4 - 570240 \zeta_5
+ 14(128354 - 722 \xi - 837 \xi^2)/3
\Big\}[16\nf] 
\nonumber\\&&
+\Big\{d_3, 1\Big\}.\Big\{1296(12 (28 - 6 \xi + \xi^2) 
- 4 (2392 + 108 \xi - 63 \xi^2 - 17 \xi^3 + 16 \xi^4) \zeta_3 
\nonumber\\&&
+ 5 (1696 + 544 \xi - 252 \xi^2 + 42 \xi^3 + 7 \xi^4) \zeta_5),
\nonumber\\&&
-4(8202784 + 512546 \xi - 111402 \xi^2 + 28107 \xi^3 - 3888 \xi^4)/3 
\nonumber\\&&
- 36 (159040 - 19104 \xi - 162 \xi^2 + 1092 \xi^3 - 123 \xi^4) \zeta_3 
+ 1296 (492 - 376 \xi + 91 \xi^2 - 9 \xi^3) \zeta_4 
\nonumber\\&&
+ 270 (28832 + 320 \xi - 732 \xi^2 + 186 \xi^3 - 7 \xi^4) \zeta_5 \Big\} \;.
\ea
We observe that \eq\nr{eq:4loop} agrees, in the SU($N$) limit and for all powers of $\xi$, with the 4-loop results of \cite{Chetyrkin:2004mf}. Furthermore, its terms of order $\xi^0$ and $\xi^1$ coincide exactly with the corresponding terms given in \cite{Czakon:2004bu}, leading us to the conclusion that it represents the correct generalization of the known results to general covariant gauge.
As a side note, we notice that the structure $\cf\,\nf^0$ is absent, and that the coefficient of $d_2\,\nf$ does not depend on the gauge parameter. In retrospect, since the same could have been observed -- at least to NLO in $\xi$ -- already in \cite{Czakon:2004bu}, assuming this pattern to hold for all powers of the gauge parameter would have allowed for a correct lift of the SU($N$) results of \cite{Chetyrkin:2004mf} to a general gauge group.

We have evaluated the five-loop contribution in Feynman gauge ($\xi=0$) as
\ba
\la{eq:gammaCc4}
2^{14}\,3^5\,\gammaCc{4} &=& \gammaCc{44}\,[16\nf]^4+\gammaCc{43}\,[16\nf]^3+\gammaCc{42}\,[16\nf]^2+\gammaCc{41}\,[16\nf]+\gammaCc{40} +\order{\xi}\;,\\
\label{eq:LO}
\gammaCc{44} &=& 3 (65 + 80 \zeta_3 - 144 \zeta_4)\;,\\
\gammaCc{43} &=& \Big\{\cf, 1\Big\}.\Big\{-2 (14765 + 12528 \zeta_3 - 38880 \zeta_4 + 20736 \zeta_5), 
\nonumber\\&&
 -3 (8325 + 15664 \zeta_3 + 12240 \zeta_4 - 33408 \zeta_5)\Big\}\;,\\
\gammaCc{42} &=& \Big\{\cf^2, \cf, d_1, d_2, 
  1\Big\}.\Big\{-72 (53927 - 182112 \zeta_3 + 48384 \zeta_3^2 + 42768 \zeta_4 + 144000 \zeta_5 - 
     86400 \zeta_6), 
\nonumber\\&&
  -4 (364361 + 484488 \zeta_3 - 1804032 \zeta_3^2 + 1868184 \zeta_4 - 2239488 \zeta_5 + 777600 \zeta_6), 
\nonumber\\&&
     20736 (107 - 109 \zeta_3 - 96 \zeta_3^2 - 36 \zeta_4 + 180 \zeta_5), 
\nonumber\\&&
     -41472 (52 \zeta_3 + 18 \zeta_3^2 - 36 \zeta_4 - 125 \zeta_5 + 
     75 \zeta_6), 
\nonumber\\&&
     2 (239495 - 3082212 \zeta_3 - 1721088 \zeta_3^2 + 3863376 \zeta_4 - 156384 \zeta_5 - 1425600 \zeta_6)\Big\}\;,\\
\gammaCc{41} &=& \Big\{\cf^3, \cf^2, \cf d_2, \cf, d_2, d_3, 
  1\Big\}.\Big\{746496 (7 + 26 \zeta_3 + 490 \zeta_5 - 560 \zeta_7), 
\nonumber\\&&
  576 (24617 - 
     301866 \zeta_3 - 196560 \zeta_3^2 + 177066 \zeta_4 + 274680 \zeta_5 - 491400 \zeta_6 + 725760 \zeta_7), 
\nonumber\\&&
     165888 (4 + 66 \zeta_3 + 216 \zeta_3^2 - 705 \zeta_5 + 357 \zeta_7), 
  16 (4796303 - 9571932 \zeta_3 + 6399648 \zeta_3^2 
\nonumber\\&&
   + 11100240 \zeta_4 - 
     16127424 \zeta_5 + 8845200 \zeta_6 - 10809288 \zeta_7), 
\nonumber\\&&
     -5184 (4192 - 
     87152 \zeta_3 + 21432 \zeta_3^2 + 5616 \zeta_4 + 89300 \zeta_5 - 27300 \zeta_6 - 
     20139 \zeta_7), 
\nonumber\\&&
  -864 (2805 - 86018 \zeta_3 - 15960 \zeta_3^2 + 43542 \zeta_4 - 70360 \zeta_5 - 68700 \zeta_6 + 192906 \zeta_7), 
\nonumber\\&&
  2 (52725013 + 136974540 \zeta_3 + 1505088 \zeta_3^2 - 118046052 \zeta_4 - 
     226012536 \zeta_5 
\nonumber\\&&
     + 84380400 \zeta_6 + 143718624 \zeta_7)\Big\}\;,\\
\la{eq:gammaCc40}
\gammaCc{40} &=& \Big\{d_3, 1\Big\}.\Big\{ -6912 (5326 + 771746 \zeta_3 - 17934 \zeta_3^2 - 209916 \zeta_4 - 
     1172870 \zeta_5 + 377625 \zeta_6 
\nonumber\\&&
        + 396669 \zeta_7), 
     -8 (192342607 + 
     174080040 \zeta_3 + 36201384 \zeta_3^2 - 103216464 \zeta_4 - 855002232 \zeta_5 
\nonumber\\&&
     + 222650100 \zeta_6 + 492202872 \zeta_7)\Big\}\;,
\ea
where \eq\nr{eq:LO} (as well as the leading-\/$\Nf$ terms at lower loops given in \eqs\nr{eq:2loop}--\nr{eq:4loop}) agrees with the respective term of the known all-loop large-\/$\Nf$ Landau-gauge expression\footnote{Note that the 1-loop fermion bubble is transverse, such that the leading-\/$\Nf$ does not pick up the gauge-parameter dependence of the bare gluon propagator. Hence, Feynman- and Landau-gauge expressions coincide.} of \cite{Gracey:1993ua}, which in full form can be written as
\ba
\gammaCC|_{\xi=1} &=& -\eta(\af)/\nf+\order{1/\nf^2} 
\;,\quad \af = 4\,a\,\nf/3
\;,\quad \eta(\ep) = \frac{(2\ep-3)\Gamma(4-2\ep)}{16\Gamma^2(2-\ep)\Gamma(3-\ep)\Gamma(\ep)} \;.
\ea
As a second check, specializing to SU(3) allows us to compare with the Feynman-gauge 5-loop result given in \eq(3.2) of \cite{Baikov:2014pja}. Again, we find full agreement.

%
\subsection{Ghost-gluon vertex anomalous dimension}

The anomalous dimension of the ghost-gluon vertex has the structure
\ba
\la{eq:gammaCCG}
\gammaCCG &=& - a (1 - \xi) \Big[
\tfrac12 
+ \tfrac{6-\xi}{8}\,a
+ \gammaCcg{2} a^2 + \gammaCcg{3} a^3 + \gammaCcg{4} a^4 + \dots \Big] \;,
\ea
where the prefactor is consistent with the known fact that the ghost vertex in Landau gauge is finite \cite{Taylor:1971ff,Blasi:1990xz} and hence does not need to be renormalized, $\gammaCCG|_{\xi=1}=0$.
The 3-loop coefficient can be found in \cite{Chetyrkin:2004mf}
\ba
2^7\,\gammaCcg{2} &=&  - 15 [16\nf] + 2 (250 - 59 \xi + 10 \xi^2) \;.
\ea
The 4-loop coefficient suffers from the same degeneracy of color factors as mentioned above, obstructing a direct  generalization from SU($N$) to a general gauge group. 
We have therefore computed it from scratch, in general covariant gauge, obtaining
\ba
\la{eq:gammaCcg3}
2^7\,3^5\,\gammaCcg{3} &=&
(-251 + 324 \zeta_3)[16\nf]^2
\nonumber\\&&
+\big(324 (96 \zeta_3 + 36 \zeta_4 - 161)\cf -6166 + 4077 \xi/2 - 
     162 (164 - 5 \xi) \zeta_3 - 8748 \zeta_4 \big) [16\nf]
\nonumber\\&&
+ 1944 \big((272 - 60 \xi + 3 \xi^2 + 7 \xi^3) \zeta_3 - 
    5 (56 - 12 \xi + 3 \xi^2 + \xi^3) \zeta_5\big)d_3
\nonumber\\&&
   + 751120 - 27 \xi (5434 - 1332 \xi + 171 \xi^2)
    +81 (2528 - 548 \xi + 99 \xi^2 - \xi^3) \zeta_3 
\nonumber\\&&
    + 1458 (4 - \xi) (2 - \xi) \zeta_4 
    - 405 (496 - 72 \xi + 9 \xi^2 + 2 \xi^3) \zeta_5  \;.
\ea
Once again, the SU($N$) limit reproduces all orders of $\xi$ as known from \cite{Chetyrkin:2004mf}, while the terms of order $\xi^0$ and $\xi^1$ coincide with those in the the linear combination $\gammaCCG=\gammaQQG-\gammaQQ+\gammaCC$ assembled from \cite{Czakon:2004bu}. As above, in retrospect, one could have observed the absence of a $\cf\,\nf^0$ term to NLO in $\xi$ from the results of \cite{Czakon:2004bu}, and by conjecturing this to hold for the full gauge-dependent result as well one could have correctly reconstructed the full 4-loop coefficient \eq\nr{eq:gammaCcg3} from the SU($N$) results of \cite{Chetyrkin:2004mf}.

At five loops, we have obtained the new Feynman gauge ($\xi=0$) result
\ba
\la{eq:gammaCcg4}
2^{14}\,3^5\,\gammaCcg{4} &=& 
\gammaCcg{43}\,[16\nf]^3+\gammaCcg{42}\,[16\nf]^2+\gammaCcg{41}\,[16\nf]+\gammaCcg{40} +\order{\xi}\;,\\
\gammaCcg{43} &=& -2989 - 1440 \zeta_3 + 5184 \zeta_4\;,\\
\gammaCcg{42} &=& \Big\{\cf, 1\Big\}.\Big\{ 1296 (557 - 736 \zeta_3 + 108 \zeta_4 + 192 \zeta_5),
\nonumber\\&&
 251891 + 1591056 \zeta_3 - 335016 \zeta_4 - 717984 \zeta_5\Big\}\;,\\
\gammaCcg{41} &=& \Big\{\cf^2, \cf, d_2, d_3, 1\Big\}.\Big\{5184 (3731 + 9588 \zeta_3 - 1440 \zeta_3^2 + 1332 \zeta_4 - 10800 \zeta_5 - 3600 \zeta_6),
\nonumber\\&&
  -1296 (45129 - 14192 \zeta_3 - 4032 \zeta_3^2 + 5616 \zeta_4 - 19296 \zeta_5 - 7200 \zeta_6), 
\nonumber\\&&
     -31104 (1360 \zeta_3 + 168 \zeta_3^2 + 144 \zeta_4 - 1260 \zeta_5 - 
     300 \zeta_6 - 441 \zeta_7), 
\nonumber\\&&
     -10368 (1126 \zeta_3 + 150 \zeta_3^2 - 567 \zeta_4 - 
     1200 \zeta_5 + 975 \zeta_6 - 441 \zeta_7), 
 - 42165410
\nonumber\\&&
 -432 (145015 \zeta_3 + 3564 \zeta_3^2 - 9168 \zeta_4 - 114001 \zeta_5 - 10950 \zeta_6 + 17640 \zeta_7)
     \Big\}\;,\\
\la{eq:gammaCcg40}
\gammaCcg{40} &=& \Big\{d_3,1\Big\}.\Big\{20736 (70330 \zeta_3 + 11076 \zeta_3^2 - 8856 \zeta_4 - 81380 \zeta_5 + 16500 \zeta_6 - 12607 \zeta_7 - 2451), 
\nonumber\\&&
  8 (114251711 + 54643392 \zeta_3 + 7060608 \zeta_3^2 - 7531704 \zeta_4 - 
     143288568 \zeta_5 + 9023400 \zeta_6 
\nonumber\\&&
     + 52599078 \zeta_7)\Big\}\;.
\ea
As an important check, we find full agreement with \eq(40) of \cite{Baikov:2015tea}, where the 5-loop term had been given for SU(3) and in Feynman gauge. Compared to the other anomalous dimensions, note that in $\gammaCCG$ there are no terms proportional to $a^\ell\,\nf^{\ell-1}$ at $\ell$ loops; these would correspond to renormalon contributions, which are absent in this case due to consistency with the vanishing of $\gammaCCG$ in Landau gauge, as has been mentioned already above.

%
\section{Complete set of renormalization constants}
\la{se:rest}

Now that the minimal set of renormalization constants is known, all remaining anomalous dimensions can be reconstructed easily, since they are related via gauge invariance of the QCD action (see e.g.\ \cite{Chetyrkin:2004mf}). From the results listed here, the anomalous dimensions of the gluon field, the gluon vertices as well as the quark-gluon vertex can be obtained from the linear relations 
\ba
\la{eq:gammaGG}
\gammaGG\;=&2(\gammaCCG-\gammaCC)-\beta \;,\quad
\gammaGGG&=\;3(\gammaCCG-\gammaCC)-\beta \;,\\
\gammaGGGG\;=&4(\gammaCCG-\gammaCC)-\beta \;,\quad
\gammaQQG&=\;\gammaCCG-\gammaCC+\gammaQQ \;,
\ea
with $\gammaQQ$ from \cite{Luthe:2016xec}. For the convenience of the reader, we attach an electronic version of the complete set of anomalous dimensions to the present paper \cite{download}.

To reconstruct the renormalization constants $Z_i$ from the set of anomalous dimensions $\gamma_i$, one starts from \eq\nr{eq:gammaDef}, recalling that in general, renormalization scale dependence enters $Z_i(a,\xi_L)$ through both of its variables. Therefore,
\ba
\la{eq:g2Z}
\gamma_i &=& 
-(\partial_{\ln\mu^2}a)(\partial_a\ln Z_i)-(\partial_{\ln\mu^2}\ln\xi_L)(\partial_{\ln\xi_L}\ln Z_i) \nonumber\\
&=& -a(\beta-\ep)(\partial_a\ln Z_i) -\gammaGG (\xi-1)(\partial_\xi\ln Z_i) \nonumber\\
&=& -a(\beta-\ep)(\partial_a\ln Z_i) -(2\gammaCCG-2\gammaCC-\beta)(\xi-1)(\partial_\xi\ln Z_i) 
\;,
\ea
where in the second line we have have been careful to use the $d$\/-dimensional version of the Beta function of \eq\nr{eq:betaDef}, exploited that the bare gauge parameter renormalizes as the gluon field $\xi_{L,b}=Z_3\xi_{L,r}$ and changed the gauge parameter to our preferred notation $\xi$ whose powers correspond to an expansion around Feynman gauge, the relation to $\xi_L$ being $\xi_L+\xi=1$.
Finally, we have for convenience traded the gluon field anomalous dimension for the ones that we have given explicitly above. 
Writing the renormalization constants as $Z_i=1+\sum_{n>0}z_i^{(n)}/\ep^n$, the coefficients $z_i^{(n)}$ then follow by solving \eq\nr{eq:g2Z}, requiring $\gammaGG$ (viz $\gammaCCG$, $\gammaCC$ and $\beta$) at one loop lower only.
Since the expressions for the complete 5-loop renormalization constants are somewhat large, we refrain from listing them here, but provide electronic versions thereof \cite{download}.

Once the RCs $Z_i$ are known, the corresponding anomalous dimensions can simply be extracted from the single poles, as $\gamma_i=a\partial_a z_i^{(1)}$.

%
\section{Conclusions}
\la{se:conclu}

In recent years, technical progress has made possible the first complete five-loop computations in non-Abelian gauge theories. Adding to the already available set of renormalization constants for general Lie groups in $\msbar$\/-like schemes \cite{Luthe:2016ima,Luthe:2016xec,Herzog:2017ohr}, we have presented new analytic 5-loop Feynman-gauge results for the two missing anomalous dimensions, which we have chosen to be those of the ghost field \eqs\nr{eq:gammaCc4}-\nr{eq:gammaCc40} and ghost-gluon vertex \eqs\nr{eq:gammaCcg4}-\nr{eq:gammaCcg40}. This completes the five-loop renormalization program for general groups, providing a generalization -- and independent confirmation -- of the previously known SU(3) coefficients \cite{Baikov:2014qja,Baikov:2014pja,Baikov:2015tea,Baikov:2016tgj}, relevant for physical QCD. 

Along the way, we have closed a gap in the literature and provided full gauge-dependent expressions for the renormalization constants of the ghost sector at four loops, see \eqs\nr{eq:4loop} and \nr{eq:gammaCcg3} above. Together with our new results, we have prepared computer-readable versions of the complete set of anomalous dimensions and renormalization constants, available online \cite{download}.

The methods we have employed here are well suited to be applied to the gluon propagator as well, and the corresponding computation of the gluon field anomalous dimension $\gamma_3$ is under way \cite{beta4paper}. The anticipated result would give an important independent check on the Feynman-gauge expression that we have provided in the ancillary files (which was derived using \eq\nr{eq:gammaGG}), and hence on the correctness of the Beta function from the independent calculation of \cite{Herzog:2017ohr}.

For completeness, it might be interesting to evaluate the gauge parameter dependence of ghost field and -vertex (as well as the quark field) in the future. While this would, for example, provide a further independent check of the correctness of the Beta function (as well as the quark mass anomalous dimension $\gamma_m$) due to gauge-parameter cancellation, given the strong constraints already discussed above we do not think this a pressing issue. However, from the viewpoint of truly completing the 5-loop renormalization program, knowledge of the full gauge dependence of all renormalization constants is certainly desirable.

In passing, we note that the analytic structure of the 5-loop Beta function (as well as that of the corresponding renormalization constant $Z_a$), containing the Zeta values $\{\zeta_3,\zeta_4,\zeta_5\}$ only, is considerably simpler than that of the other anomalous dimensions and RCs, which in addition need the weight-6 and weight-7 constants $\{\zeta_3^2,\zeta_6,\zeta_7\}$. Regarding the group-theoretic structure, the only outlier is $\gammaCCG$ (and $Z_1^{ccg}$), where the factor $d_1$ is absent.

%
\acknowledgments

The work of T.L.\ has been supported in part by DFG grants GRK 881 and SCHR 993/2; he is grateful to the theory group of the University of Bielefeld for hospitality, and for continued access to parts of their computer clusters.
A.M.\ is supported by a European Union COFUND/Durham Junior Research Fellowship under EU grant agreement number 267209.
P.M.\ was supported in part by the EU Network HIGGSTOOLS PITN-GA-2012-316704.
Y.S.\ acknowledges support from FONDECYT project 1151281 and UBB project GI-152609/VC.

%
\appendix
\section{Beta function}
\label{se:beta}

Being an important fundamental parameter of gauge theories, considerable effort has gone into evaluating the Beta function over the past four decades. After groundbreaking work at one loop \cite{Gross:1973id,Politzer:1973fx}, establishing the asymptotically free nature of the strong coupling constant, the 2-loop \cite{Caswell:1974gg,Jones:1974mm}, 3-loop \cite{Tarasov:1980au,Larin:1993tp} and 4-loop \cite{vanRitbergen:1997va,Czakon:2004bu} perturbative corrections have been evaluated, made possible by major technological developments that were pushed ahead in parallel. Five-loop results have appeared over the past 8 years or so, first for the case of QED \cite{Baikov:2008cp,Baikov:2010je,Baikov:2012zm}, later for SU(3) \cite{Baikov:2016tgj,Chetyrkin:2016uhw}, and finally for general Lie groups \cite{Luthe:2016ima,Herzog:2017ohr}.

In order to translate the recent results of \cite{Herzog:2017ohr} to our notation (as introduced in \se\ref{se:notation} and \eq\nr{eq:a}), we define the $L$\/-loop coefficients $b_{L-1}$ of the Beta function as
\ba
\partial_{\ln\mu^2}\,a=-a\Big[\ep-\beta\Big]=-a\Big[\ep+b_0\,a+b_1\,a^2+b_2\,a^3+b_3\,a^4+b_4\,a^5+\dots\Big]
\;.
\ea
The coefficients are polynomials in $\nf$, and up to four loops read
\ba
3^1\,b_0 &=& \big[\!-4\big]\nf+11 \;,\\
3^2\,b_1 &=& \big[\!-36\cf-60\big]\nf+102\;,\\
3^3\,b_2&=& \big[132\cf+158\big]\nf^2+\big[54\cf^2-615\cf-1415\big]\nf+2857/2\;,\\
3^5\,b_3&=& \big[1232\cf+424\big]\nf^3
+432(132\zeta_3-5)d_3
+(150653/2-1188\zeta_3)
+\\\nonumber
&& \big[72(169-264\zeta_3)\cf^2+64(268+189\zeta_3)\cf
+1728(24\zeta_3-11)d_1
+6(3965+1008\zeta_3)
\big]\nf^2
+\\\nonumber
&& \big[11178\cf^3+\!36(264\zeta_3\!-\!1051)\cf^2+\!(7073\!-\!17712\zeta_3)\cf 
+\!3456(4\!-\!39\zeta_3)d_2
+\!3(3672\zeta_3\!-\!39143)
\big]\nf,
\ea
which the five-loop coefficient can be represented as
\ba
\la{eq:result}
3^5\,b_4 &=&b_{44}\,\nf^4+b_{43}\,\nf^3+b_{42}\,\nf^2+b_{41}\,\nf+b_{40} \;,\\
b_{44} &=& \Big\{\cf,1\Big\}.\Big\{-8(107+144\zeta_3),4(229-480\zeta_3)\Big\} \;,\\
b_{43} &=& \Big\{\cf^2,\cf,d_1,1\Big\}.\Big\{
-6(4961-11424\zeta_3+4752\zeta_4),
-48(46+1065\zeta_3-378\zeta_4),
\nonumber\\&&
1728(55-123\zeta_3+36\zeta_4+60\zeta_5), 
-3(6231+9736\zeta_3-3024\zeta_4-2880\zeta_5)
\Big\} \;,\\
b_{42} &=& \Big\{\cf^3, \cf^2, \cf d_1, \cf, d_2, d_1, 1\Big\}. \Big\{
-54 (2509 + 3216 \zeta_3 - 6960 \zeta_5), 
\nonumber\\&&
9(94749/2 - 28628 \zeta_3 + 10296 \zeta_4 - 39600 \zeta_5), 
25920 (13 + 16 \zeta_3 - 40 \zeta_5), 
\nonumber\\&&
3(5701/2 + 79356 \zeta_3 - 25488 \zeta_4 + 43200 \zeta_5), 
-864 (115 - 1255 \zeta_3 + 234 \zeta_4 + 40 \zeta_5), 
\nonumber\\&&
-432 (1347 - 2521 \zeta_3 + 396 \zeta_4 - 140 \zeta_5), 
843067/2 + 166014 \zeta_3 - 8424 \zeta_4 - 178200 \zeta_5\Big\} \;,\\
b_{41} &=& \Big\{\cf^4, \cf^3, \cf^2, \cf d_2, \cf, d_3, d_2, 1\Big\}. \Big\{
-81(4157/2 + 384 \zeta_3), 
81 (11151 + 5696 \zeta_3 - 7480 \zeta_5), 
\nonumber\\&&
-3 (548732 + 151743 \zeta_3 + 13068 \zeta_4 - 346140 \zeta_5), 
-25920 (3 - 4 \zeta_3 - 20 \zeta_5), 
\\&&
8141995/8 + 35478 \zeta_3 + 73062 \zeta_4 - 706320 \zeta_5, 
216 (113 - 2594 \zeta_3 + 396 \zeta_4 + 500 \zeta_5), 
\nonumber\\&&
216 (1414 - 15967 \zeta_3 + 2574 \zeta_4 + 8440 \zeta_5), 
-5048959/4 + 31515 \zeta_3 - 47223 \zeta_4 + 298890 \zeta_5\Big\} \;,
\nonumber\\
b_{40} &=& \Big\{d_3,1\Big\}. \Big\{-162 (257-9358 \zeta_3+1452 \zeta_4+7700 \zeta_5),
\nonumber\\&&
8296235/16 - 4890 \zeta_3 + 9801 \zeta_4/2 - 28215 \zeta_5\Big\} \;.
\ea
Of these 5-loops coefficients, $b_{44}$ has been known already for a long time from a large-\/$\Nf$ analysis \cite{PalanquesMestre:1983zy,Gracey:1996he}, while $b_{43}$ was given in \cite{Luthe:2016ima}, as a proof-of-concept of our setup that we have used above to determine the anomalous dimensions of the ghost sector. 
The three coefficients $b_{42}$, $b_{41}$ and $b_{40}$ have recently been derived as well \cite{Herzog:2017ohr}, using the background field method and relying on infrared rearrangement \cite{Vladimirov:1977ak} and the $R^*$ operation \cite{Chetyrkin:1984xa} to map the ultraviolet divergences onto massless four-loop two-point functions which were evaluated via their new code \code{FORCER} \cite{Ueda:2016yjm}.
As an important check on the 5-loop expressions given above, setting the group invariants to their SU(3) values \eq\nr{eq:su3}, all coefficients coincide with the results given in \cite{Baikov:2016tgj}.

%

\end{document}